\documentclass[pra,twocolumn,superscriptaddress]{revtex4}

\usepackage{times,amsmath,amsfonts,amssymb,latexsym,textcomp}
\usepackage{graphicx}
\usepackage{epstopdf}
\usepackage{float}

\usepackage[usenames]{color}
\usepackage{epsf}

\newcommand{\ket}[1]{\left\vert#1\right\rangle}
\newcommand{\bra}[1]{\left\langle#1\right\vert}

\renewcommand{\phi}{\varphi}
\renewcommand{\epsilon}{\varepsilon}
\newcommand{\eff}{\textrm{eff}}

\DeclareFontFamily{U}{euc}{}
\DeclareFontShape{U}{euc}{m}{n}{<-6>eurm5<6-8>eurm7<8->eurm10}{}
\DeclareSymbolFont{AMSc}{U}{euc}{m}{n}
\DeclareMathSymbol{\umu}{\mathord}{AMSc}{"16}

\begin{document}
\title{Observation of topologically protected bound states \\ in a one dimensional photonic system}

\author{Takuya Kitagawa$^{1\dagger}$}
\author{Matthew A. Broome$^{3 \dagger}$}
\author{Alessandro Fedrizzi$^3$}
\author{Mark S. Rudner$^1$}
\author{Erez Berg$^1$} 
\author{Ivan Kassal$^2$}
\author{Al\'an Aspuru-Guzik$^2$} 
\author{Eugene Demler$^1$}
\author{Andrew G. White$^3$}
\affiliation{$^1$Department of Physics and $^{2}$Department of Chemistry and Chemical Biology, Harvard University, Cambridge MA 02138, United States, $^3$ARC Centre for Engineered Quantum Systems and ARC Centre for Quantum Computation and Communication Technology, School of Mathematics and Physics, University of Queensland, Brisbane 4072, Australia}
\affiliation{$\dagger$These authors contributed equally to this work.}

\date{\normalsize{\today}}

\maketitle

\begin{figure*}[th]
\begin{center}
\includegraphics[width = 15cm]{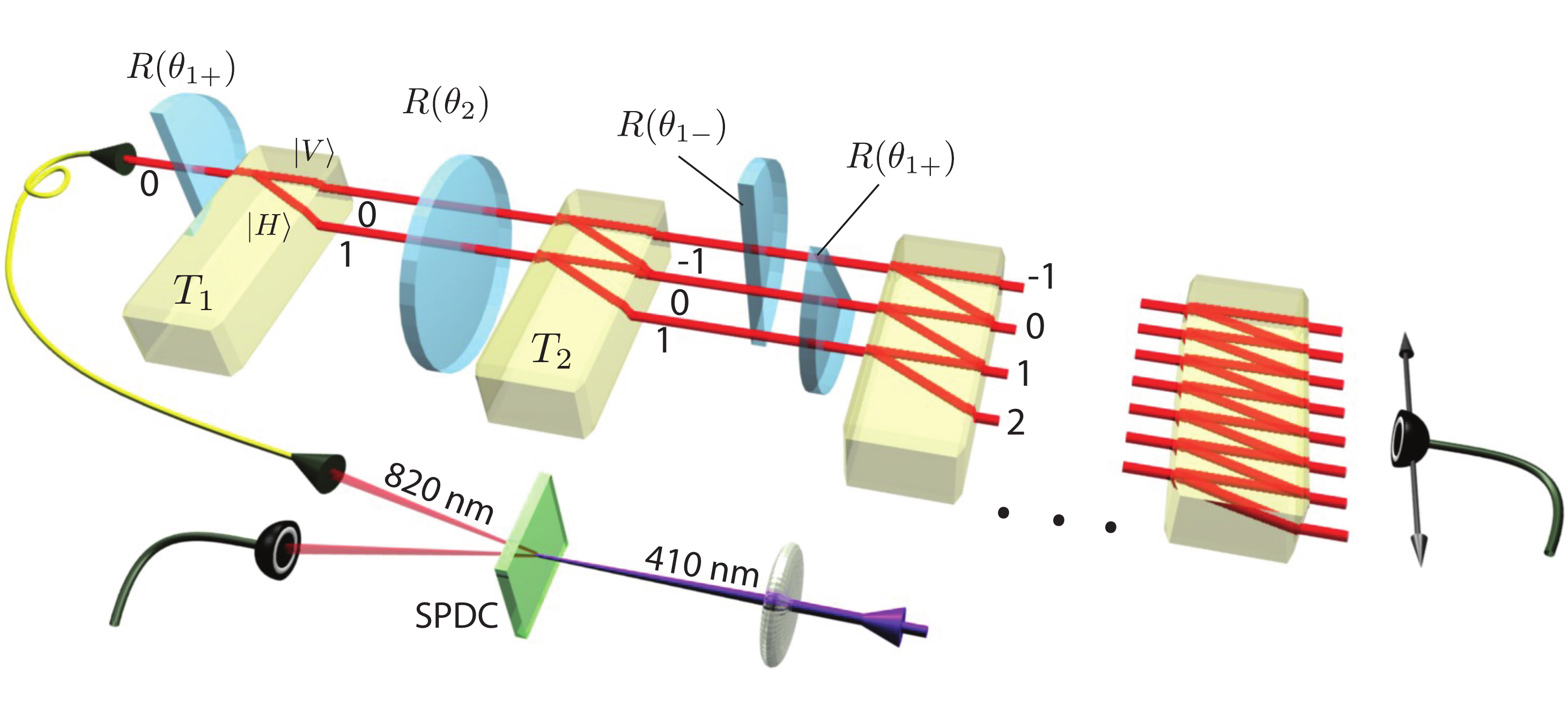}
\caption{Split-step quantum walk experiment. A polarization-encoded single photon, created via spontaneous parametric downconversion (SPDC), undergoes a succession of steps consisting of rotations, $R(\theta_1), R(\theta_2)$ and translations, $T_1$, $T_2$, implemented by half-wave plates and birefringent beam displacers respectively~\cite{UQ2010}. 
In the split-step protocol~\cite{PhysRevA.82.033429}, the two translations displace first $\ket{H}$ and then $\ket{V}$; experimentally, we implement such translations through birefringent beam displacers and by shifting the lattice origin by $+1$ site after each full step. To probe the topological properties of the quantum walk, semi-circular half-wave plates are used to create spatially inhomogeneous rotations, $R(\theta_{1-})$, $R(\theta_{1+})$. The output probability distribution is imaged with a single-photon avalanche detector.}
\label{fig:splitstep}
\end{center}
\end{figure*}

{\bf
One of the most striking features of quantum mechanics is the appearance of phases of matter with topological origins.  These phases result in remarkably robust macroscopic phenomena such as the edge modes in integer quantum Hall systems~\cite{PhysRevLett.45.494}, the gapless surface states of topological insulators~\cite{Chen2009,Xia2009}, and elementary excitations with non-abelian statistics in fractional quantum Hall systems and topological superconductors\cite{RevModPhys.80.1083}. Many of these states hold promise in the applications to quantum memories and quantum computation\cite{RevModPhys.80.1083,  PhysRevLett.100.096407, 2010arXiv1006.4395A,Wray:2010fk, Kitaev20032}. 
Artificial quantum systems, with their precise controllability, provide a versatile platform for creating and probing a wide variety of topological phases\cite{Wang2009,PhysRevA.82.033429,sorensen,zhu,jaksch, lewenstein}. Here we investigate topological phenomena in one dimension, using photonic quantum walks\cite{PhysRevA.82.033429}.
The photon evolution simulates the dynamics of topological phases which have been predicted to arise in, for example, polyacetylene. We experimentally confirm the long-standing prediction of topologically protected localized states associated with these phases by directly imaging their wavefunctions. Moreover, we reveal an entirely new topological phenomenon: the existence of a topologically protected pair of bound states which is unique to periodically driven systems\cite{PhysRevB.82.235114}. Our experiment demonstrates a powerful new approach for controlling topological properties of quantum systems through periodic driving.}

The distinguishing feature of topological phases is the existence of a winding in the ground state wave function of the system, which cannot be undone by gentle changes to the microscopic details of the system.
Such topological structures appear in a variety of physical contexts, from condensed-matter~\cite{Chen2009,Xia2009,RevModPhys.80.1083, PhysRevLett.100.096407, 2010arXiv1006.4395A, Wray:2010fk,RevModPhys.82.3045,2010arXiv1008.2026Q} and high-energy physics~\cite{PhysRevD.13.3398} to quantum optics~\cite{Wang2009} and atomic physics~\cite{PhysRevA.82.033429,sorensen,zhu,jaksch,lewenstein}. These systems provide diverse platforms for studying the universal features of topological phases and their potential for technological applications. 

In this paper we study topological phenomena in periodically driven systems using the discrete time quantum walk~\cite{aharonov_quantum_1993}, a protocol for controlling the motion of quantum particles on a lattice. In particular, we demonstrate that quantum walks stroboscopically simulate topological phases~\cite{PhysRevA.82.033429} which belong to the same topological class as that of the Su-Schrieffer-Heeger (SSH) model of polyacetylene~\cite{PhysRevLett.42.1698} and the Jackiw-Rebbi model of a one-dimensional spinless Fermi field coupled to a Bose field~\cite{PhysRevD.13.3398}. These two models, developed in entirely different fields, share a common underlying topological structure, which has
been predicted to result in the existence of topologically protected bound states with exactly zero energy. Intriguingly, such zero-energy bound states are responsible for the existence of solitons with fractional fermion number in both models~\cite{Jackiw1981253}. However, to date such zero-energy bound states have never been directly observed. In this experiment, we confirm the existence of these topologically robust bound states 
for the first time by directly imaging their wavefunctions.

An additional advantage of investigating topological phenomena with quantum walks is that this apparatus 
allows us to access the dynamics of strongly driven systems far from the static or adiabatic regime~\cite{PhysRevB.82.235114,Jiang2011,lindner2011fti}, to which most previous work on topological phases has been restricted. In this strongly-driven regime, we discover a topologically protected pair of \emph{non-degenerate} bound states. This phenomenon is unique to periodically driven systems and has not been reported before.

\begin{figure*}
\begin{center}
\includegraphics[width = 15cm]{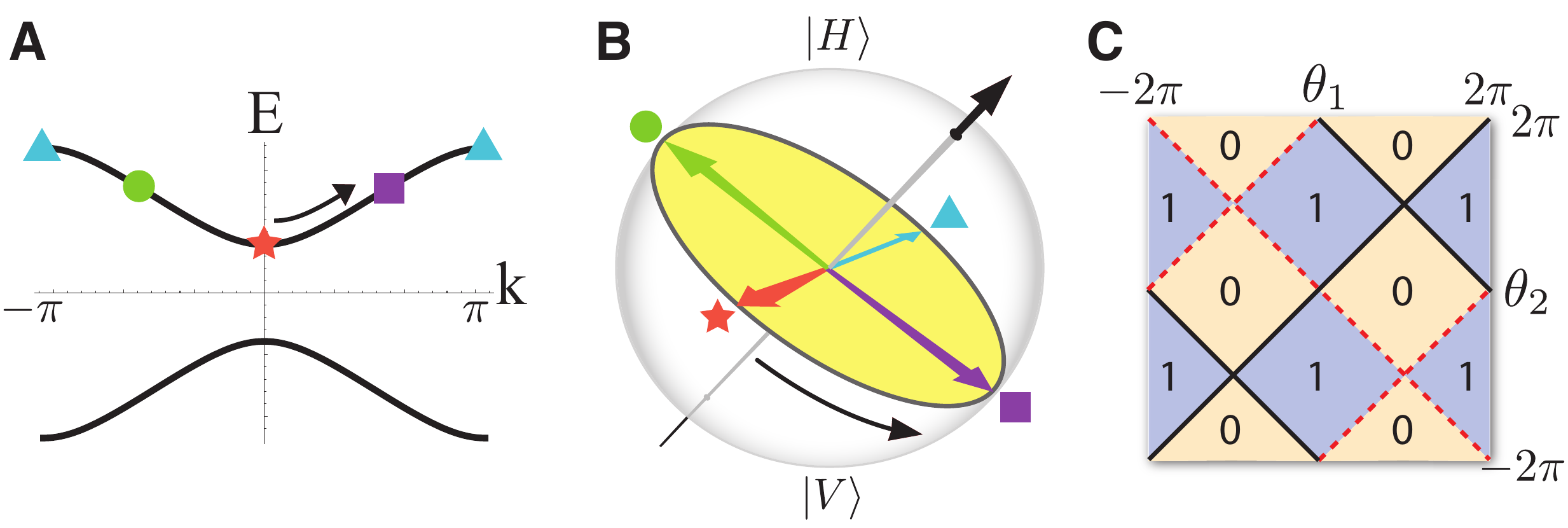}
\caption{{\bf A,} A typical band structure of the effective Hamiltonian $H_{\textrm{eff}}(\theta_{1},\theta_{2})$ for the split-step
quantum walk (Here, $\theta_{1}{=}\pi/2$ and $\theta_{2}{=}0$). The two bands 
correspond to the eigenvalues of $H_{\textrm{eff}}(\theta_{1},\theta_{2})$. For most $\theta_{1}$ and $\theta_{2}$, the
bands display a gap. 
{\bf B,} Topology of $H_{\textrm{eff}}(\theta_{1},\theta_{2})$. Each eigenstate of $H_{\textrm{eff}}(\theta_{1},\theta_{2})$ with momentum $k$ corresponds to a point on a Bloch sphere, illustrated by the symbols in A. 
As $k$ runs from $-\pi$ to $\pi$, the states follow a 
closed trajectory around a great circle, and the winding number $W$ characterizes the topology of $H_{\textrm{eff}}(\theta_{1},\theta_{2})$. 
{\bf C,} Phase diagram of $H_{\textrm{eff}}(\theta_{1},\theta_{2})$ which shows the winding number $W$ as a function of $\theta_1$ and $\theta_2$. The transition lines correspond to points where the spectral gap closes at eigenvalues $E{=}0$ (black solid line) and $E=\pi$ (red dotted line).}
\label{fig:band}
\end{center}
\end{figure*}

Discrete time quantum walks have been realized in several physical architectures~\cite{MichalKarski07102009, PhysRevLett.104.100503, PhysRevLett.103.090504, PhysRevLett.104.050502,UQ2010}. 
Here we use the photonic setup demonstrated in Ref.~\onlinecite{UQ2010} to implement a variation of these walks, the split-step quantum walk~\cite{PhysRevA.82.033429} of a single photon with two internal states on a one-dimensional lattice, see Fig.~\ref{fig:splitstep}.
The two internal states are encoded in the horizontal, $\ket{H}$, and vertical, $\ket{V}$, polarization components of the photon. One step of the split-step quantum walk consists of i) a polarization rotation $R(\theta_1)$
followed by a polarization-dependent translation $T_1$ of $\ket{H}$ to the right by one lattice site, and ii) a second rotation $R(\theta_2)$, followed by the translation $T_2$, of $\ket{V}$ to the left, see Appendix \ref{app:1}. 
The quantum walk is generated by repeated applications of the one-step operator $U(\theta_1, \theta_2){=}T _{2}R(\theta_2)T_{1} R(\theta_1)$. 
 In the paraxial approximation, the propagation of the photon in the static experimental setup in Fig.~\ref{fig:splitstep} is described by an effective time-dependent Schr\"{o}dinger equation with periodic driving that corresponds to this quantum walk.


The topological structure underlying split-step quantum walks is revealed by studying the effective Hamiltonian $H_{\textrm{eff}}(\theta_1, \theta_2)$, defined through $U(\theta_1, \theta_2){=} e^{-iH_\eff(\theta_1, \theta_2)}$. 
In a stroboscopic sense, the discrete-time dynamics of the quantum walk are equivalent to dynamics generated by 
$H_\eff(\theta_1, \theta_2)$ viewed after unit time intervals.
In this way, the quantum walk simulates $H_\eff(\theta_1, \theta_2)$. 
A typical spectrum of $H_\eff(\theta_1, \theta_2)$ is gapped, as illustrated in Fig.~\ref{fig:band}A. 
The non-trivial topological structure of $H_\eff(\theta_1, \theta_2)$ is due to a chiral symmetry~\cite{PhysRevA.82.033429}. In translationally invariant systems with this symmetry, the polarization of an eigenstate of $H_\eff(\theta_1, \theta_2)$ with momentum $k$, when represented as a spinor on the Bloch sphere, follows a path along a great circle as the momentum $k$ goes from $-\pi$ to $\pi$ (see Fig.~\ref{fig:band}B and Appendix \ref{app:2}). The topology is then characterized by the winding number $W$ of this path around the origin. For the split-step quantum walk, two distinct phases with $W{=}0$ and $W{=}1$ exist, see 
Fig.~\ref{fig:band}C.

\begin{figure*}[!th]
\begin{center}
\includegraphics[width = 13cm]{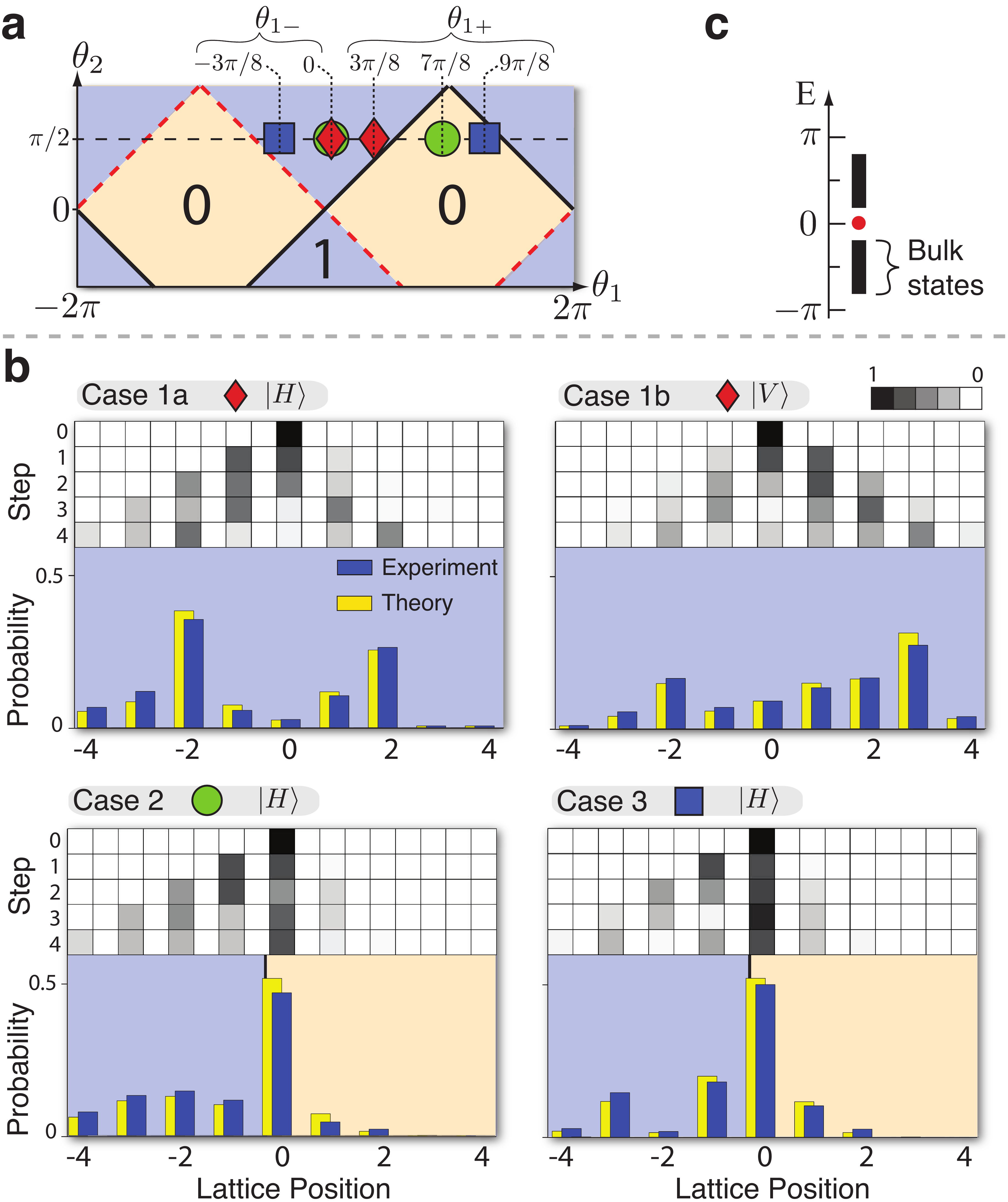}
\caption{
{\bf a,} Phase diagram, with symbols indicating the parameters ($\theta_{1-}$, $\theta_{1+}$, $\theta_{2}$), 
and the winding numbers for each experimental case.
{\bf b,} Experimental probability distributions, with $\theta_2{=}\pi/2$.  
The rotation $R(\theta_{1})$ is spatially inhomogeneous with $\theta_{1-}$ ($\theta_{1+}$)
in the region $x{<}0$ ($x{\ge}0$). The results show the absence of a bound state near $x{=}0$ 
for both initial photon polarizations $\ket{H}$ and $\ket{V}$ in cases 1a and 1b, respectively. 
Case 2 shows the presence of a bound state with a pronounced peak near $x{=}0$ after 4 steps. 
The bar graphs compare the measured (blue) and predicted (yellow) probabilities after the fourth step. 
Case 3 demonstrates that the presence of the bound state is robust against changes of parameters.
Experimental errors due to photon counting statistics are not visible on this scale.  
{\bf c,} Calculated quasi-energy spectrum of the effective Hamiltonian for case 2. The bound state with quasi-energy $E{=}0$ (red dot) is analogous to the zero-energy states of the SSH and Jackiw-Rebbi models. 
}
\label{fig:result1}
\end{center}
\end{figure*}

A striking consequence of non-trivial topology is the appearance of localized states at boundaries between two topologically distinct phases~\cite{shinsei,PhysRevD.13.3398,PhysRevA.82.033429,PhysRevLett.42.1698}. 
Because our experimental setup allows access to individual lattice sites, we are able to probe this phenomenon by creating 
a boundary between regions where dynamics are governed by two gapped Hamiltonians $H_\eff(\theta_{1-}, \theta_2)$ and $H_\eff(\theta_{1+}, \theta_2)$ characterized by winding numbers $W_-$ and $W_+$. 
Here we choose to create the boundary by making $\theta_1$ inhomogeneous with $\theta_1(x){=}\theta_{1-}$ for lattice positions $x{<}0$ and $\theta_1 (x){=}\theta_{1+}$ for $x{\ge}0$. 
When $W_-\neq W_+$, it is expected that topologically robust localized states exist at the boundary near $x=0$.
This can be understood in a heuristic fashion as follows. 
When $W_-\neq W_+$, the winding number $W_-$ of the bulk gapped Hamiltonian $H_\eff(\theta_{1-}, \theta_2)$ can only be changed to that of $H_\eff(\theta_{1+}, \theta_2)$ given by $W_+$ by closing the gap of the system,
see Fig.~\ref{fig:band}C. 
Thus, near the boundary at $x=0$ between these two regions, the energy gap closes, and it is expected that 
states exist within the gaps of the bulk spectra of $H_\eff(\theta_{1-}, \theta_2)$ and  $H_\eff(\theta_{1+}, \theta_2)$. 
Because extended states do not exist in this energy range, such a state is necessarily localized at the boundary.
This argument shows that a change in topology at a boundary is accompanied by the presence of a localized state.
Therefore, due to the topological origin of these localized states, they are robust against perturbations\cite{shinsei} such as small changes of quantum walk parameters or the presence of a static disordered potential caused by, for example, small spatial variations of rotation angles $\theta_{1}$ and $\theta_{2}$. 

\begin{figure*}[!th]
\begin{center}
\includegraphics[width =13cm]{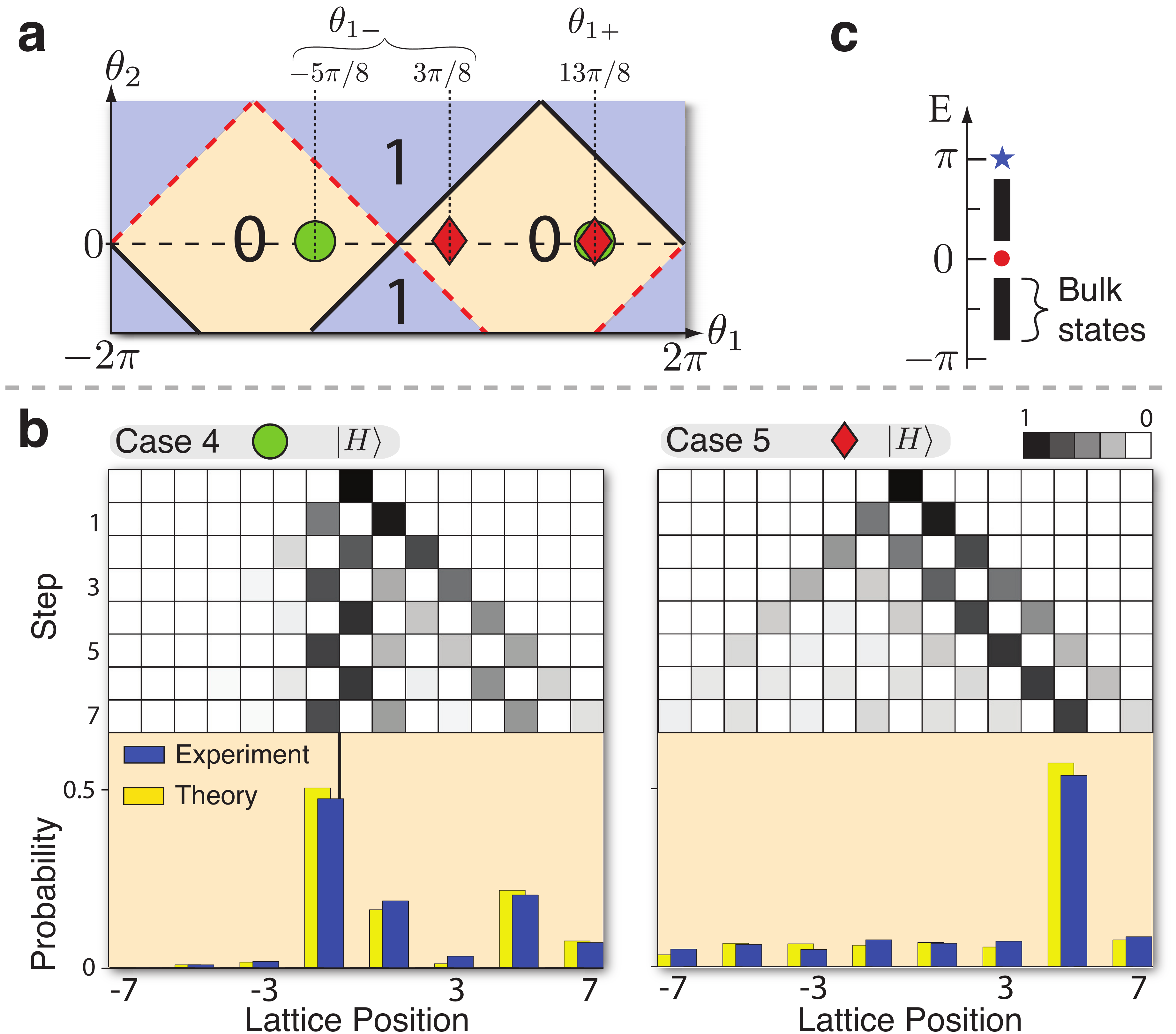}
\caption{
{\bf a,} Phase diagram, with symbols indicating the parameters ($\theta_{1-}$, $\theta_{1+}$, $\theta_{2}$), and winding numbers for each experimental case.
{\bf b,} Experimental probability distributions, with $\theta_2 {=}0$. In case 4 we observe oscillatory probabilities around $x{=}0$, indicating the presence of at least two bound states with the quasi-energy difference of $\pi$. They are absent in case 5 
for initial polarization of $\ket{H}$. Other initial polarizations and parameters have been implemented and the result is presented in SI. The bar graphs compare the measured (blue) and predicted (yellow) probabilities after the seventh step.
{\bf c,} Quasi-energy spectrum of case 4. In addition to the $E{=}0$ (red dot) bound state, there is a $E{=}\pi$ bound state (blue star), whose topological origin is described in the text and SI. }
\label{fig:result2}
\end{center}
\end{figure*}

To probe the existence of the bound states, we initialize a photon next to the boundary 
between two topologically distinct quantum walks, Fig.~\ref{fig:splitstep}. In the absence of bound states, the photon is expected to spread ballistically, with the detection probability at the origin quickly decreasing to zero. However, if there is a bound state, the bound state component of the initial state will remain near this boundary even after many steps.

We first implemented split-step quantum walks with $\theta_{2}{=}\pi/2$ and $\theta_{1-}$ and $\theta_{1+}$ such that $W_-{=}W_+{=}1$, shown as case 1 on the phase diagram in Fig.~\ref{fig:result1}a. 
In both cases 1a and 1b in Fig.~\ref{fig:result1}b with the initial polarization of $\ket{H}$ and $\ket{V}$, respectively, 
the detection probability  at the origin quickly decreases to zero. On the other hand, for case 2 in Fig.~\ref{fig:result1}b with 
parameters chosen to create a boundary between topologically distinct phases $W_-{=}1$ and $W_+{=}0$, 
we observe the existence of at least one bound state as a peak in the probability distribution near the origin after four steps. This boundary state is a direct analogue of the zero-energy states of the SSH and Jackiw-Rebbi models~\cite{PhysRevLett.42.1698, PhysRevD.13.3398}. The quasi-energy $E$ of the localized state, i.e. the eigenvalue of the effective Hamiltonian associated with this state, can be found by explicit calculation, see Fig.~\ref{fig:result1}c.
Here we indeed find a single state at $E=0$.
The versatile control over parameters in our experimental apparatus allows the test of the robustness 
of these states against a variety of changes in microscopic parameters, which is a universal 
feature of topological states~\cite{PhysRevLett.42.1698, PhysRevD.13.3398}. 
To test this, we implemented case 3  where $\theta_{1-}$ and $\theta_{1+}$ are shifted from those of case 2 while maintaining $W_-{=}1$ and $W_+{=}0$, and confirmed the existence of a bound state in Fig.~\ref{fig:result1}b.
In addition, we study the effects of controlled amounts of decoherence on the bound states
and present the result in Appendix \ref{app:3}.

Our experiment also reveals a new topological phenomenon unique to periodically driven systems, which can be probed by studying split-step quantum walks with $\theta_2{=}0$, see Fig.~\ref{fig:result2}.
With the appropriate choice of basis (see Appendix \ref{app:4}), this quantum walk becomes equivalent to the one described by
the one-step operator $U{=}i T R(\theta_1)$, where $T{=}T_1 T_2$ can be implemented with a single beam displacer, extending the experiment to seven steps.
This class of quantum walks can only realize a single topological phase characterized by 
the winding number $W{=}0$. Therefore we do not expect bound states for spatially inhomogeneous $\theta_{1}$ based on winding numbers.
However, the evolution of the probability distribution for case 4 displays period-2 oscillations in the vicinity of the origin. 
This observation strongly suggests the existence of at least two bound states 
whose quasi-energies differ by $\pi$. 
In the Appendix \ref{app:5} we show that this pair of bound states with quasi-energy difference $\pi$ is robust against small changes of $\theta_{1-}$ and $\theta_{1+}$.
On the other hand, in case 5 we demonstrate that such bound states are absent when $\theta_{1-}$ and $\theta_{1+}$ are continuously connected without crossing topological phase boundaries, see Fig.~\ref{fig:result2}. 

The existence of this pair of bound states with quasi-energy difference $\pi$ is a robust phenomenon 
with new topological origin which has not been studied previously. 
Chiral symmetry implies eigenstates of a Hamiltonian generally come in pairs with
quasi-energies $E$ and $-E$. 
In the case of a static Hamiltonian, the symmetry makes a zero-energy state special since this energy satisfies $E=-E$,
and therefore a single state at $E=0$ is topologically protected~\cite{shinsei}. For a periodically-driven system, because 
the effective Hamiltonian is defined through a one-step evolution operator by $U= e^{-iH_\eff(\theta_{1}, \theta_2)}$, 
the quasi-energies of $H_\eff(\theta_{1}, \theta_2)$ are defined only up to $2\pi$. In particular, $E=\pi$ and $E=-\pi$
correspond to the same quasi-energy, and therefore
$E=\pi$ represents another special value of quasi-energy satisfying $E=-E$. Thus, like
zero-energy state of static systems, a single $\pi$ quasi-energy state is topologically protected~\cite{PhysRevB.82.235114, Jiang2011}. The coexistence of such $E=0$ and $E=\pi$ states suggested by the period-2 oscillations observed in case 4 is
checked through the explicit calculation of quasi-energy spectrum presented in Figure~\ref{fig:result2}C
for case 4. In the Appendix \ref{app:6} and \ref{app:7}, we give the characterization of this structure in terms of topological invariants of periodically
driven systems and demonstrate their topological robustness. 


Our work opens up a rich arena for future research.
First, the experiment demonstrates the direct imaging of bound states, providing a powerful tool for probing a variety of topological phenomena. 
Second, the versatility of our setup allows
for extensions, such as the realization of other topological phases in one and two dimensions~\cite{PhysRevA.82.033429}, 
the study of many-photon quantum walks with non-linear interactions, as well as the exploration of 
new topological phenomena unique to periodically driven systems in higher dimensions~\cite{PhysRevB.82.235114}. 

We thank B. P. Lanyon, B. J. Powell for discussions. We acknowledge financial support from the ARC Centres of Excellence, Discovery and Fed. Fellow programs and an IARPA-funded US Army Research Office contract. 
T. K., M. S. R., E. R. and E. D. thank DARPA OLE program,
CUA, NSF under DMR-07-05472, AFOSR Quantum Simulation MURI, and the ARO-MURI on Atomtronics.
I. K. and A. A.-G. thank the Dreyfus and Sloan Foundations, ARO under
W911-NF-07-0304 and DARPA's Young Faculty Award N66001-09-1-2101-DOD35CAP.

\newpage
\appendix

 \begin{center}
    {\bf APPENDICES}
  \end{center}

\section{Rotation operators implemented in the experiment} \label{app:1}
The implementations of split-step quantum walks with a photon require the rotations of polarizations, 
written as $R(\theta)$ in the main text. 
In this experiment, we used half-wave plates
which implements $R(\theta){=}e^{-i \sigma_{y} \theta/2} \sigma_{z} $, where $\sigma_{i}$ are Pauli matrices
such that $\sigma_{z} \ket{H} =  \ket{H}$ and $\sigma_{z} \ket{V} =  -\ket{V}$. 

\section{Winding numbers of split-step quantum walk}\label{app:2}

Ref. \cite{PhysRevA.82.033429} considered creating a boundary between regions with different topological numbers by varying the second rotation angle $\theta_{2}$. We described
the topological structure of the split-step quantum walk in terms of the one-step evolution operator, or Floquet operator, $U(\theta_{1}, \theta_{2}) = T_{2} R(\theta_{2}) T_{1} R(\theta_{1})$ 
and associated chiral symmetry operator $\Gamma_{\theta_1}$, which depends only on $\theta_1$ and satisfies  $\Gamma_{\theta_1}^{-1} U(\theta_1, \theta_2) \Gamma_{\theta_1}=U^{\dagger}(\theta_1, \theta_2)$. 
In this experiment, we implemented inhomogeneous split-step quantum walks by varying the first rotation angle, $\theta_{1}$. In order to maintain the chiral symmetry in the system, it is necessary to characterize the dynamics in terms of an alternative chiral symmetry operator that depends only the second rotation angle, $\theta_{2}$. 
In the following, we explain and define such a chiral symmetry operator. 
As a consequence of considering such a chiral symmetry operator, 
the phase diagrams in the main text are slightly different from those in Ref. \cite{PhysRevA.82.033429}.

\begin{figure*}[!th]
\centering
\includegraphics[width=15cm]{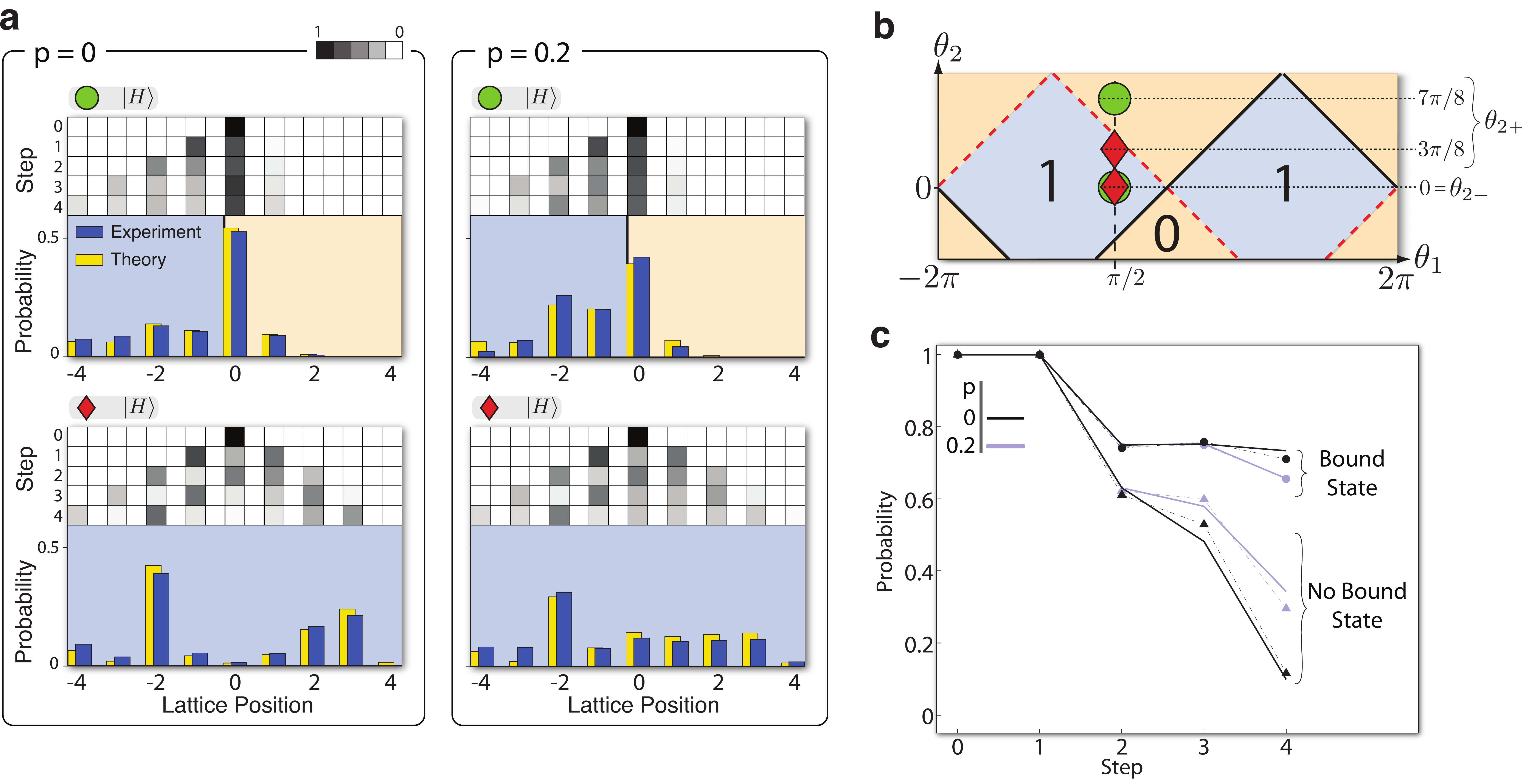}
\caption{
{\bf a,} Probability distributions for a split-step quantum walks with and without additional decoherence and initial state $\ket{H}$.
In this case, in contrast to the results in Fig.3 in the main text, $\theta_{1}{=}\pi/2$ and $\theta_{2}$ is made inhomogeneous. 
{\bf b,} Phase diagram, with symbols indicating the parameters  ($\theta_{1-}$, $\theta_{1+}$, $\theta_{2}$), and the topological phases for each case. 
{\bf c,} Sum of probabilities at lattice positions around the boundary ($-1$, $0$ and $+1$) for integer steps of the split-step quantum walk. The solid lines show theoretical predictions and the dashed lines are the experimental results, error bars are smaller than the marker size. The difference between bound and unbound states can be seen despite the introduction of decoherence into the system.
\label{fig:deco1} }
\end{figure*}

Because the origin of time for a periodically driven system is arbitrary, 
we can characterize the topology of the split-step quantum walk with a different 
initial time, namely in terms of the evolution operator 
$U'(\theta_{2}, \theta_{1}) = T_{1} R(\theta_{1}) T_{2} R(\theta_{2})$. 
This alternative choice corresponds to making a half-period shift of the origin of time.
Using the momentum-space expressions $T_{1} = \sum_{k} e^{ik \sigma_{z}/2} e^{ik/2} \ket{k}\bra{k} $
and $T_{2} = \sum_{k} e^{ik \sigma_{z}/2} e^{-ik/2} \ket{k}\bra{k} $, we see that
$U'(\theta_{2}, \theta_{1})$ is different from $U(\theta_{1}, \theta_{2})$ only through the exchange
of $\theta_{1}$ and $\theta_{2}$,  i.e. $U'(\theta_{2}, \theta_{1}) = U(\theta_{2}, \theta_{1})$.  
Therefore, it is clear that the chiral symmetry operator of $U'(\theta_{2}, \theta_{1})$ is given by $\Gamma_{\theta_{2}}$, and that the winding numbers of $U'(\theta_{2}, \theta_{1})$ are the same as 
those of $U(\theta_{1}, \theta_{2})$.
The chiral symmetry of $U'(\theta_{2}, \theta_{1})$ 
only depends on the second rotation angle $\theta_{2}$, and thus the symmetry is preserved  
even when $\theta_{1}$ is varied in space. 
Therefore it is possible to construct inhomogeneous quantum walks with boundaries between topologically distinct phases, while preserving the required chiral symmetry across the entire system.

\begin{figure*}[!th]
\centering
\includegraphics[width=15cm]{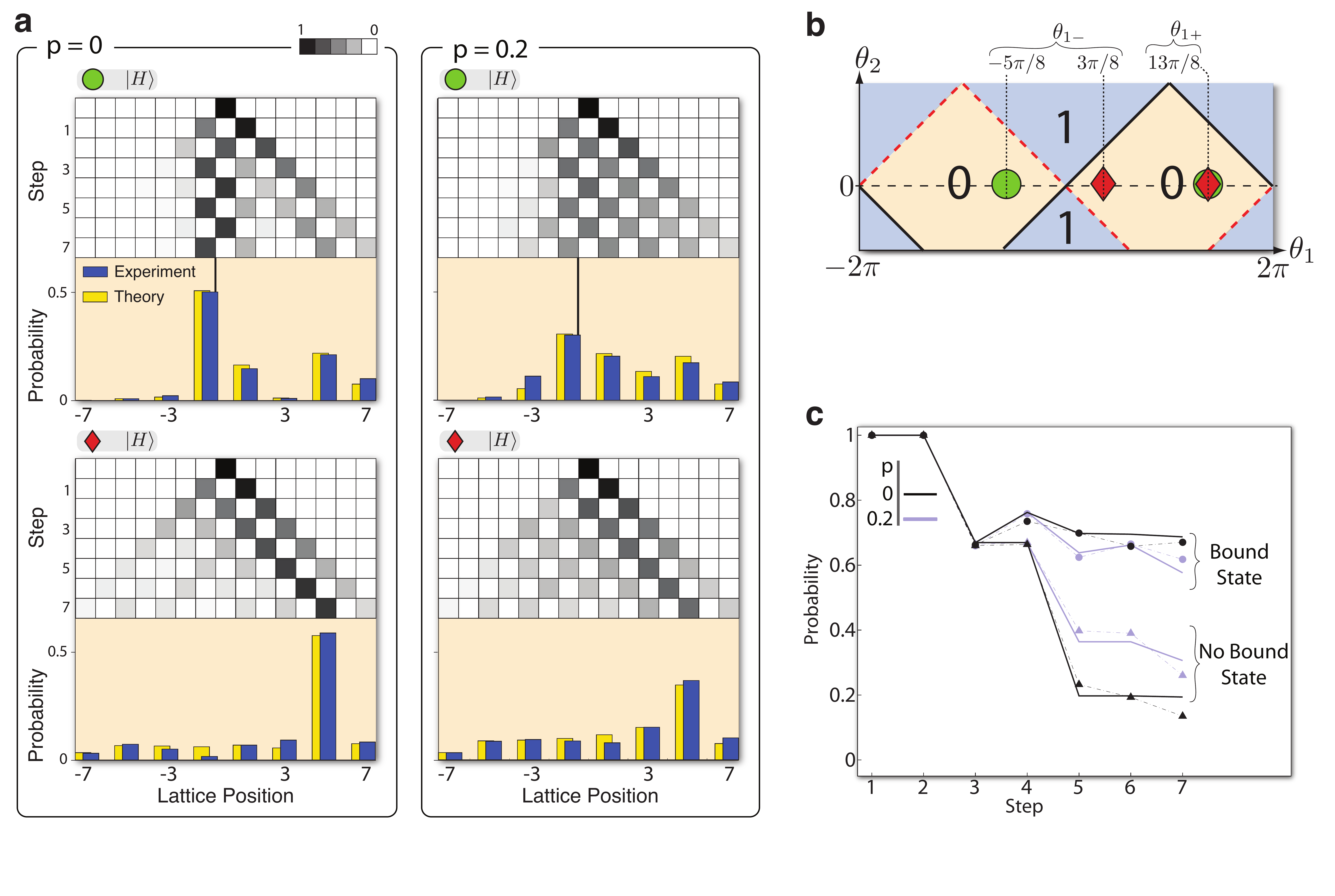}
\caption{
{\bf a,} Effect of decoherence on the pair of bound states in split-step quantum walk c.f. cases $4$ and $5$,
with $\theta_{2}{=}0$ and initial state $\ket{H}$.
{\bf b,} Phase diagram indicating the topological phases for each case.
{\bf c,} Sum of probabilities at lattice positions around the boundary ($-1$, $0$ and $+1$) for integer steps of the split-step quantum walk. The solid lines show theoretical predictions and the dashed lines are the experimental results, error bars are smaller than the marker size. The difference between bound and unbound states can be seen despite the introduction of decoherence into the system.
\label{fig:deco2} }
\end{figure*}

\section{Bound states under decoherence}\label{app:3}
In this supplementary information, we present the result of quantum walks in the presence of controlled amounts of 
dephasing. 
While the topological bound states observed in the
paper are no longer  stationary states of the evolution under dephasing, signatures of 
such bound states are observable for a small number of steps as we show in this Supplementary Material. 
This result demonstrates that it is possible to study topological phenomena, for short time dynamics, 
in other systems that might be more prone to decoherence. 

One feature of our optical quantum walk setup is the ability to tune the level of decoherence~\cite{UQ2010}. Each pair of beam displacers forms an interferometer, which can be intentionally misaligned to add temporal and spatial walkoffs~\cite{UQ2010}. This process, coupled with measurement of the photon, 
and corresponds well to pure dephasing~\cite{UQ2010}. If the system at step $N$ is described by the density matrix $\rho_{N}$ it will evolve according to:
\begin{equation}
\rho_{N+1}=(1-p)U\rho_{N}U^\dagger + p\sum_i K_{i}U\rho_{N}U^\dagger K^\dagger_{i}
\label{eq:decoherence}
\end{equation}
where $p$ is the amount of dephasing and $K_{i}$ are the associated Kraus operators. For $p{=}0$, Eq.~\ref{eq:decoherence} describes a pure quantum walk, while $p{=}1$ represents a system without any quantum coherence, {\it i.e.}
the evolution is described by classical random walks.

Figure~\ref{fig:deco1} shows the results for the split-step quantum walks with and without additional dephasing, corresponding to $p{=}0.2$ and $p{=}0$, respectively.
In this case the rotation $R(\theta_{2})$ is inhomogeneous and $\theta_{1}{=}\pi/2$. 
Accordingly, we define the winding number corresponding to the chiral operator $\Gamma_{\theta_{1}}$ in 
Figure~\ref{fig:deco1}, see Supplementary Section 1 for details. 
Note that $p=0$ indicates that we do not introduce additional dephasings, but this case still could contain 
decoherence coming from experimental limitations. 
The rotation angles $\theta_{1}$ and $\theta_{2}$ studied in this experiments are indicated in the phase diagram Fig.~\ref{fig:deco1}B.
In the presence of dephasing, the bound state observed in Fig.~\ref{fig:deco1}A (top) gradually decays as the number of 
steps increases. However, for a small amount of dephasing, this decay is slow, and for a small number of steps, 
the probability distribution is still sharply peaked near the boundary compared to the cases with no bound states,
as displayed in Fig.~\ref{fig:deco1}A (bottom).
In addition, Fig.~\ref{fig:deco1}C quantifies the effect of decoherence on the probability distribution around the boundary. 

In addition, we studied the effect decoherence on on cases $4$ and $5$ discussed in the main text and 
present the result in Figure~\ref{fig:deco2}A, B and C. Again, for small amount of decoherence, the signature of 
bound states is still observable for a small number of steps. 

\begin{figure*}[!th]
\begin{center}
\includegraphics[width = 13cm]{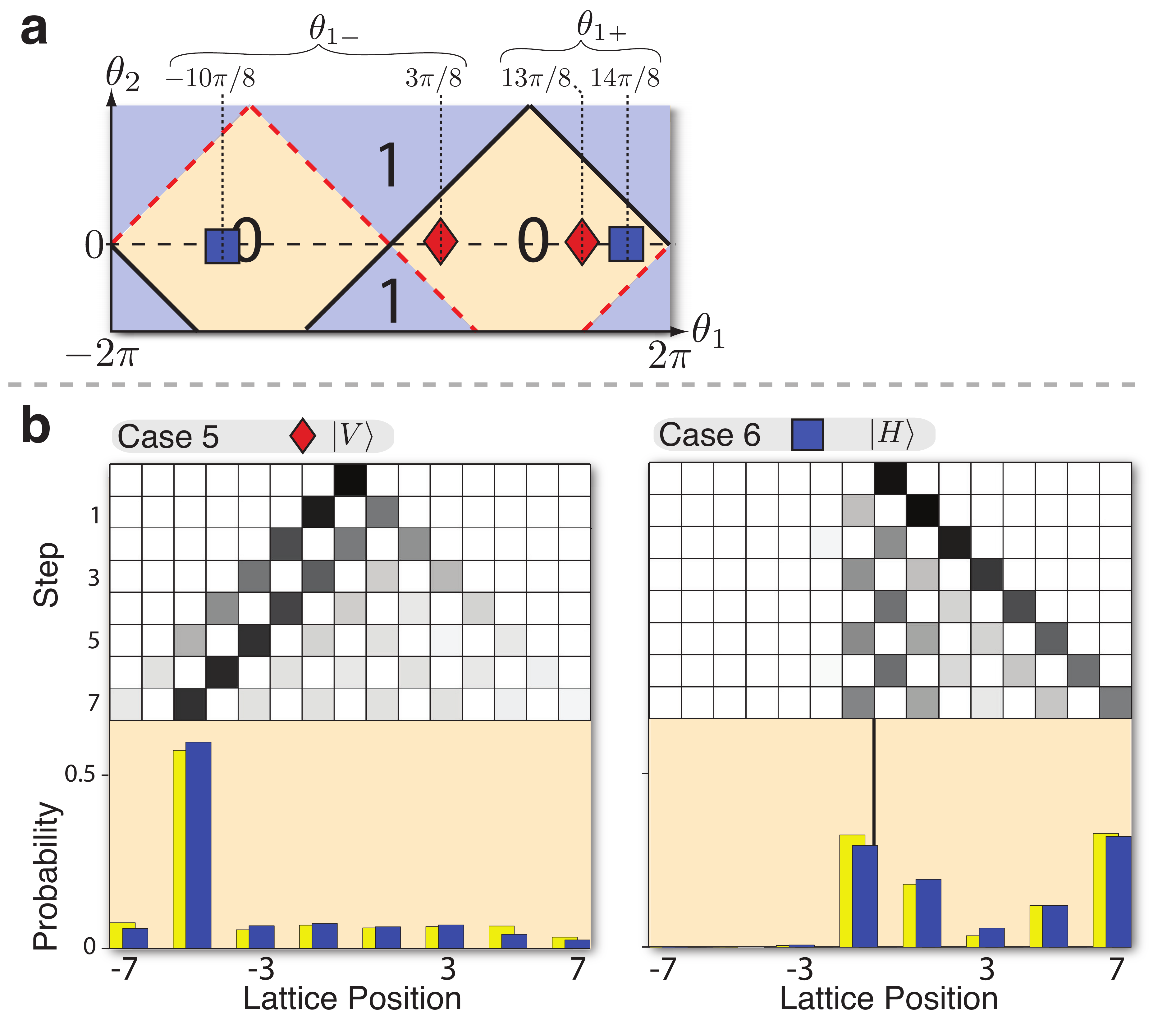}
\caption{
{\bf (A)} Phase diagram, with symbols  indicating the parameters ($\theta_{1-}$, $\theta_{1+}$, $\theta_{2}$), and the topological phases for each experimental case.
{\bf (B)} Case 5, for initial polarization state $\ket{V}$ in conjunction with case 5, initial polarization $\ket{H}$, in the main text, conclusively proves the absence of bound states for this choice of parameters. Case 6 demonstrates the topological robustness of the pairs of bound states observed in case 4: They are still visible around the origin despite a change in $\theta_{1-}$, $\theta_{1+}$. The bar graphs compare the measured (blue) and predicted (yellow) probabilities after the seventh step.}
\label{fig:floquetpair}
\end{center}
\end{figure*}

\section{The split-step quantum walk with $\theta_{2}=0$}\label{app:4}

In the main text, we studied the behavior of the split-step quantum walk 
$U=T_{2} R(\theta_{2})T_{1} R(\theta_{1})$
with $\theta_{2}=0$. Note that $R(\theta{=}0){=}\sigma_{z}$.
In the experiment, we implemented the quantum walk 
with Floquet operator $U_{ex}=T_{2}T_{1} R(\theta_{1})$. In this section, we show that these 
two quantum walks are related through a unitary transformation, and therefore
represent an equivalent dynamics of the system. In particular, the topological 
properties of these two dynamics are equivalent. 

The split-step quantum walk with $\theta_{2}=0$ is described by the 
Floquet operator $U(\theta_1,0) = T_{2} \sigma_{z} T_{1} e^{-i\sigma_{y} \theta_{1}/2} \sigma_{z}
= T_{2}T_{1}   e^{i\sigma_{y} \theta_{1}/2} $. 
It is simple to check that the unitary transformation $V= e^{-i\hat{x}\pi/2}$ acts on $T=T_{2}T_{1}$ such
that $V^{-1} T V = i T \sigma_{z}$ where $\hat{x}$ is the coordinate operator. Therefore, $U(\theta_1,0)$ and $U_{ex}$ are 
unitarily related through $V^{-1} U(\theta_1,0) V = i T  e^{-i\sigma_{y} \theta_{1}/2} \sigma_{z} = iU_{ex}$, as we 
claimed in the text. Apart from a global phase, the experimental 
implementation is equivalent to the split-step quantum walk with 
$\theta_{2}=0$.

\section{Absence and presence of a pair of topologically protected bound states}\label{app:5}
Here we provide additional experimental data, supporting the absence and presence of 
a pair of topologically protected bound states presented in the main text. 

Case 5 shown in Fig.~\ref{fig:floquetpair} corresponds to the same parameters as case 5 in the main text, and here we have 
implemented the experiments with different initial polarizations. In the main text, the initial polarization
was $\ket{H}$ whereas here we present the result for $\ket{V}$. For either initial state, we find the absence
of any bound state, and detection probability quickly decreases to zero near the boundary $x=0$. 
Because the states $\ket{H}$ and $\ket{V}$ span the space of internal states of the walker, this shows that indeed there is no bound state near $x = 0$.

Case 6, on the other hand, tests the robustness of the pair of bound states found in case 4. 
Parameters of case 6 are chosen such that $\theta_{1-}$ and $\theta_{1+}$ are both continuously 
connected with those of case 4 without crossing gapless phases in the phase diagram. Indeed, 
we observe the period-2 oscillations in the evolution of probability distributions just as in case 4, indicating the existence 
of a pair of bound states whose quasi-energies differ by $\pi$.

\section{Robustness of a single $0$ and $\pi$ energy state}\label{app:6}
Here we explain the robustness of a localized, single $0$ and $\pi$ energy state
against perturbations observed in case 4 in Figure 4 in the main text through a simple argument. 

Chiral symmetry is defined by the existence of an operator $\Gamma$ with action 
on the Hamiltonian $\Gamma H \Gamma^{-1} = -H$. Consequently, a state $\ket{\psi}$ with
energy $E$ implies the existence of a state $\Gamma^{-1} \ket{\psi}$ with energy $-E$, as can be easily
checked. Thus, in the presence of chiral symmetry, a state with energy $E$ necessarily comes in a pair
with a state with energy $-E$, except for the special case $E=-E$.
This argument applies equally to static Hamiltonians and effective Hamiltonians of periodically driven
systems, where the eigenstates of static Hamiltonian are replaced by eigenstates of effective Hamiltonians,
or Floquet states, and energies of static Hamiltonians are replaced by quasi-energies of effective Hamiltonians. 

For quasi-energy $E=0$ and $E=\pi$ for the effective Hamiltonian of periodically driven systems, 
a single state can exist at these quasi-energies.
Moreover, such a state cannot be removed or shifted in quasi-energy by weak, symmetry-preserving perturbations, because
a single state cannot be split into two.
Consequently, the localized states at quasi-energy $E=0$ and $E=\pi$ cannot be removed 
unless the bulk band gap of the effective Hamiltonian closes at $E{=}0$ and/or $E{=}\pi$, 
allowing the hybridization with the extended bulk states. 

\section{Topological invariants associated with bound states at quasi-energies $0$ and $\pi$ } \label{app:7}
In this section, we show that the topological classification of the periodically driven systems with chiral symmetry
is given by $Z \times Z$, and give the explicit expression of the topological invariants
 in terms of the wave functions of the bound states. This gives yet another understanding 
 of the topological protection of $0$ and $\pi$ energy bound states found in the experiment. 
 
 In the following, we consider the bound states at energy $0$ (analogous arguments apply to the bound states at $\pi$). 
 Suppose that there are $N_0$ degenerate bound states with energy 0, which we label $\ket{\phi_{\alpha'}^{0}}$ with $\alpha' =1 \cdots N_{0}$.
 Let the chiral symmetry of the system to be $\Gamma$, which anticommutes with the Hamiltonian, $\{ \Gamma, H\}=0$. As a consequence, $\Gamma^2$ commutes with $H$.  
 When there is no conserved quantity associated with $\Gamma^2$ \cite{Ryu2010}, it is possible to 
 choose the phase of $\Gamma$ such that $\Gamma^2=1$. For example, in the case of 
 the split-step quantum walk, we choose $\Gamma = i\Gamma_{\theta_{1}}$. 
 Because by definition $\Gamma \ket{\phi_{\alpha'}^{0}}$ is an eigenstate of $H$ with energy $0$, we can choose 
 the basis of zero energy states such that they are eigenstates of $\Gamma$. We denote 
 the zero energy states in this basis as  $\{\ket{\psi_{\alpha}^{0}}\}$ and their eigenvalues under $\Gamma$ as 
 $\{Q^0_{\alpha}\}$. 
 Since $\Gamma^2=1$, $Q^0_{\alpha}$ is either $\pm 1$. 

We now show that 
 the sum of eigenvalues, $Q^{0} \equiv \sum_\alpha Q^0_\alpha$, represents the topological invariant associated with 
 zero energy bound states. 
We define the integer $Q^{0}$ for zero-energy bound states
 and $Q^{\pi}$ for $\pi$-energy bound states constructed in an analogous fashion, as 
 \begin{eqnarray}
 Q^{0} = \sum_{\alpha} \bra{\psi_{\alpha}^0} \Gamma \ket{\psi_{\alpha}^0}  \nonumber \\
  Q^{\pi} = \sum_{\alpha} \bra{\psi_{\alpha}^{\pi}} \Gamma \ket{\psi_{\alpha}^{\pi}}  \nonumber
  \end{eqnarray}
  where $\{ \ket{\psi_{\alpha}^{\pi}} \}$ are the $\pi$ energy bound states. 
  In order to show that these quantities are indeed topological invariants, we show that perturbations of 
  the Hamiltonian which preserve the chiral symmetry cannot mix the zero- and $\pi$-energy bound states with the same eigenvalues of $\Gamma$, and therefore cannot change the energies of these states away from $0$ or $\pi$.
  Let $H'$ be a perturbation to the system such that  $\{ \Gamma, H'\}=0$. Now we evaluate the matrix 
  element of $\{ \Gamma, H'\}=0$ in the $0$ ($\pi$) energy states. The result is 
  \begin{eqnarray}
  0 &=&  \bra{\psi_{\alpha}^0} \{ \Gamma, H'\} \ket{\psi_{\beta}^0} \nonumber \\
  &=&\left\{ \begin{array}{ll} 
  2  \bra{\psi_{\alpha}^0} H'\ket{\psi_{\beta}^0} \quad \textrm{for} \quad Q_{\alpha} = Q_{\beta} \nonumber \\
  \bra{\psi_{\alpha}^0} H'\ket{\psi_{\beta}^0} -  \bra{\psi_{\alpha}^0} H'\ket{\psi_{\beta}^0}  =0\end{array} \right. \nonumber \\
 &&  \quad \quad \quad \quad  \quad \quad \quad \quad \quad \textrm{for} \quad Q_{\alpha} \neq Q_{\beta}  \nonumber
  \end{eqnarray}
  Thus, in accordance with degenerate perturbation theory, bound states with the same eigenvalues $Q_{\alpha}$ 
  cannot mix, while those with different eigenvalues in general do mix and are not protected by chiral symmetry. 
Because one can break up any finite change of the Hamiltonian into successive
  changes of small perturbations, one can repeat this argument
  and show that the values $Q^{0}$ and $Q^{\pi}$ cannot change unless 
  the bound states at $0$ and $\pi$ energies mix with the bulk states. 

  In the simple limiting case of the split-step quantum walk that we considered in Supplementary Section 3, with
  $\theta_{2}=0, \theta_{1-}=-\pi, \theta_{1}=\pi$, we can analyze the bound states of 
  the shifted evolution operator $U'(\theta_{2}, \theta_{1})= T_{1} R(\theta_{1}) T_{2}$ with
  chiral operator $\Gamma_{\theta_{2}} =\sigma_{x}$. 
  The bound-state wavefunctions can be easily computed in this limit, and one finds 
  the zero-energy bound state is associated with $Q^{0} = 1$ and $\pi$ energy bound state
  is associated with $Q^{\pi} = -1$. Because the pair of bound states found in the experiment arises in a situation which is 
  continuously connected with this special split-step quantum walk without closing the gaps, the observed pair is characterized by the same values of the topological invariants.

\bibliography{topo_qw_nature}
\bibliographystyle{naturemag}

\end{document}